\documentstyle[preprint,prl,aps]{revtex}

\begin{document}

\bibliographystyle{prsty}

\tighten

\title{Quantum filter for non-local polarization properties of\\
photonic qubits}

\author{Holger F. Hofmann and Shigeki Takeuchi}
\address{CREST, Japan Science and Technology Corporation (JST),\\ 
Research Institute for Electronic Science, Hokkaido 
University,\\
Kita-12 Nishi-6, Kita-ku, Sapporo 060-0812, Japan}

% \date{\today}

\maketitle

\begin{abstract}
We present an optical filter that
transmits photon pairs only if they share the
same horizontal or vertical polarization,
without decreasing the quantum coherence 
between these two possibilities. Various
applications for entanglement manipulations 
and multi-photon qubits are discussed.
\end{abstract}

%\pacs{PACS numbers:
%42.50.-p  % Quantum Optics
%03.67.-a  % Quantum Information
%03.65.Ud  % Entanglement and nonlocality}
\vspace{0.5cm}

The manipulation and control of polarization entanglement
between individual photons is an important challenge
for optical implementations of quantum information concepts.
Recently, significant progress has been achieved by
proposals for phase gates using only linear optical 
elements, single photon sources, and highly efficient 
detectors \cite{Kni01,Ral01}. These proposals show that strong
non-linear effects can be implemented by exploiting 
post-selection strategies based on recently developed
single photon technologies \cite{Tak99,Kim99,Lou00,Mic00,San01}. 
In the following, we apply
the same principles to realize a filter sensitive only to
the relative polarization of two photons. Since this filter
is not sensitive to the individual polarization of the 
photons, it can preserve and even create entanglement
between the photons passed through it. The function
of this filter is one of the most elementary operations on
the four dimensional Hilbert space of a pair of photonic 
qubits. It can thus be applied to a wide range of 
problems, from quantum 
non-demolition measurements of entanglement to the generation
of multi-photon quantum codes. 

The basic element of our filter is a simple beam splitter
of reflectivity $R=1/2$. The operation of this beam splitter
can be described by a unitary transformation $\hat{U}_{1/2}$
acting on the joint quantum state of the two input modes.
We now consider only cases where one of the two modes has
a one photon input and a one photon output. This case can be
realized by using a single photon source to supply one
photon coinciding with the input field state in the other
mode, and by post-selecting the one photon output events
on the same side of the beam splitter with a highly efficient
photon detector. The operation of this device on the input
state is then described by the operator $\hat{S}_{11}$ obtained
from the appropriate matrix elements of the beam splitter
operator $\hat{U}_{1/2}$,
\begin{equation}
\langle m;1 \!\mid\! \hat{U}_{1/2} \!\mid\! n;1\rangle
= \langle m \!\mid\! \hat{S}_{11} \!\mid\! n\rangle =
\delta_{n,m} \left(\frac{i}{\sqrt{2}}\right)^{n+1} (n-1).
\end{equation}
In particular, note that $\hat{S}_{11}\mid\! 1\rangle=0$,
meaning that this operation of the beam splitter eliminates
the one photon component of the input state. On the other
hand, photon pairs are transmitted since 
$\hat{S}_{11}\mid\! 2\rangle\neq 0$. 
By using the post-selection condition, we have thus turned the
simple beam splitter into a non-linear filter transmitting 
photons only if they come in pairs. 

This filter can now be applied to photon pairs in orthogonal
input modes by using the photon bunching properties of
the $R=1/2$ beam splitter, specifically
\begin{eqnarray}
&&\hat{U}_{1/2} \!\mid\! 1;1\rangle =
\frac{i}{\sqrt{2}}(\mid\! 0;2\rangle+\mid\! 2;0\rangle)
\nonumber \\ &&
\hat{U}_{1/2} 
\frac{i}{\sqrt{2}}(\mid\! 0;2\rangle+\mid\! 2;0\rangle)
= - \mid\! 1;1\rangle.
\end{eqnarray}
These two transformations show that
photon pairs entering a Mach-Zehnder interferometer setup
from different sides of the input beam splitter will always 
be found in the same path of the interferometer, but
will separate again at the output. By using the non-linear
filter on both paths of a Mach-Zehnder setup, it is therefore
possible to transmit photons only if one photon enters
from each side of the input beam splitter.  
The operator for this two mode transformation reads
\begin{eqnarray}
\label{eq:core}
\hat{U}_{1/2}(1,2)
\;(\hat{S}_{11}(1)\otimes\hat{S}_{11}(2))\; \hat{U}_{1/2}(1,2) 
\mid\! 1;1 \rangle &=& -\frac{1}{4} \mid\! 1;1 \rangle 
\nonumber \\
\hat{U}_{1/2}(1,2)
\;(\hat{S}_{11}(1)\otimes\hat{S}_{11}(2))\; \hat{U}_{1/2}(1,2) 
\mid\! 0;1 \rangle &=& \hspace{0.4cm} 0
\nonumber \\[0.1cm]
\hat{U}_{1/2}(1,2)
\;(\hat{S}_{11}(1)\otimes\hat{S}_{11}(2))\; \hat{U}_{1/2}(1,2) 
\mid\! 1;0 \rangle &=& \hspace{0.4cm} 0 
\nonumber \\
\hat{U}_{1/2}(1,2)
\;(\hat{S}_{11}(1)\otimes\hat{S}_{11}(2))\; \hat{U}_{1/2}(1,2) 
\mid\! 0;0 \rangle &=& \hspace{0.4cm}
\frac{1}{2} \mid\! 0;0 \rangle.
\end{eqnarray}
Note that the two photon component is both phase shifted and
attenuated with respect to the vacuum component. The phase
shift is easily compensated by a linear element operating 
on either one of the output fields. The attenuation can be
compensated when the filter is applied to the polarization
of a two photon input. 

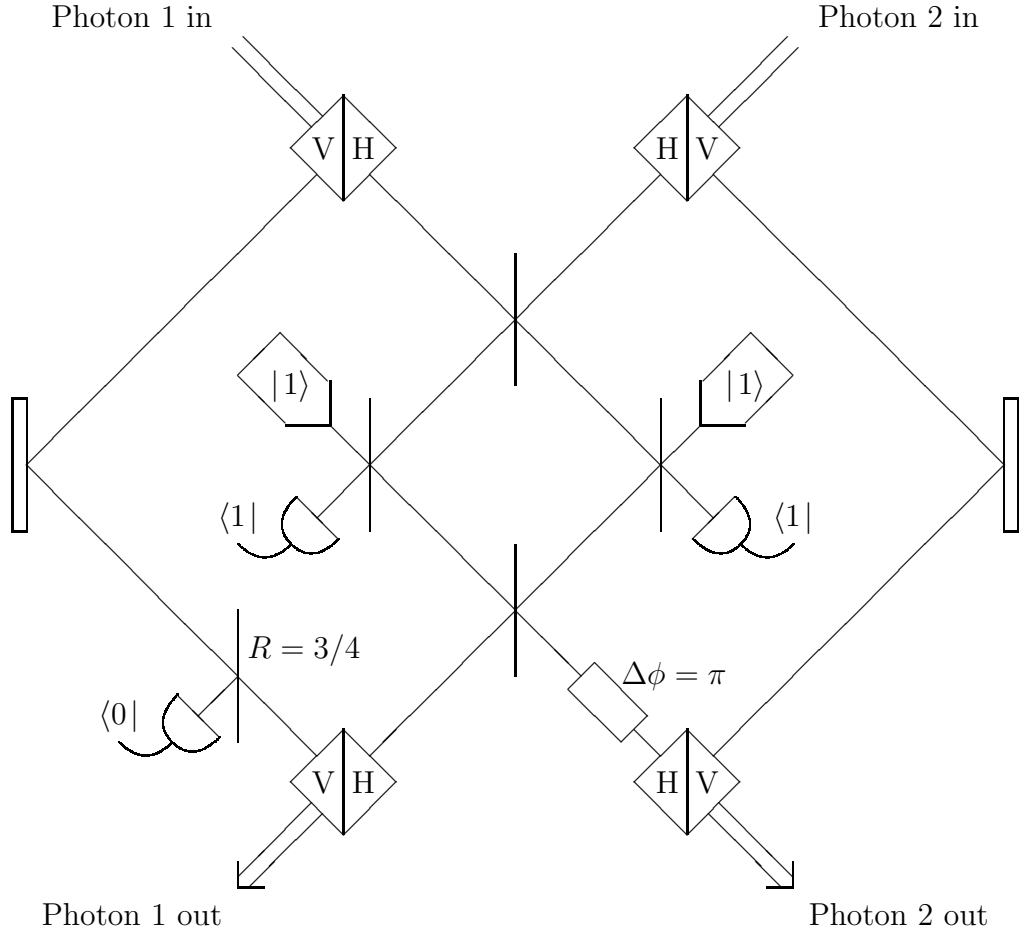
\begin{figure}
\begin{picture}(400,400)
%\put(0,0){\framebox(400,400){}}

%====V1

%--photon1in
\put(123,328){\line(-1,1){30}}
\put(127,332){\line(-1,1){30}}
\put(15,360){\makebox(80,20){Photon 1 in}}
%--polarizer1in
\put(135,300){\line(0,1){40}}
\put(135,340){\line(1,-1){20}}
\put(135,340){\line(-1,-1){20}}
\put(135,300){\line(1,1){20}}
\put(135,300){\line(-1,1){20}}
\put(120,310){\makebox(15,20){V}}
\put(135,310){\makebox(15,20){H}}
%--V1 path
\put(15,200){\line(1,-1){110}}
\put(10,175){\framebox(5,50){}}
\put(15,200){\line(1,1){110}}
%--polarizer1out
\put(135,60){\line(0,1){40}}
\put(135,100){\line(1,-1){20}}
\put(135,100){\line(-1,-1){20}}
\put(135,60){\line(1,1){20}}
\put(135,60){\line(-1,1){20}}
\put(120,70){\makebox(15,20){V}}
\put(135,70){\makebox(15,20){H}}
%--photon1out
\put(123,72){\line(-1,-1){28}}
\put(127,68){\line(-1,-1){28}}
\put(95,40){\line(1,0){10}}
\put(95,40){\line(0,1){10}}
\put(15,20){\makebox(80,20){Photon 1 out}}

%====V2

%--photon2in
\put(277,328){\line(1,1){30}}
\put(273,332){\line(1,1){30}}
\put(305,360){\makebox(80,20){Photon 2 in}}
%--polarizer2in
\put(265,300){\line(0,1){40}}
\put(265,340){\line(-1,-1){20}}
\put(265,340){\line(1,-1){20}}
\put(265,300){\line(-1,1){20}}
\put(265,300){\line(1,1){20}}
\put(250,310){\makebox(15,20){H}}
\put(265,310){\makebox(15,20){V}}
%--V2 path
\put(385,200){\line(-1,-1){110}}
\put(385,175){\framebox(5,50){}}
\put(385,200){\line(-1,1){110}}
%--polarizer2out
\put(265,60){\line(0,1){40}}
\put(265,100){\line(-1,-1){20}}
\put(265,100){\line(1,-1){20}}
\put(265,60){\line(-1,1){20}}
\put(265,60){\line(1,1){20}}
\put(250,70){\makebox(15,20){H}}
\put(265,70){\makebox(15,20){V}}
%--photon2out
\put(277,72){\line(1,-1){28}}
\put(273,68){\line(1,-1){28}}
\put(305,40){\line(-1,0){10}}
\put(305,40){\line(0,1){10}}
\put(305,20){\makebox(80,20){Photon 2 out}}

%====H12

%--upper and lower beam splitters
\put(200,230){\line(0,1){50}}
\put(200,120){\line(0,1){50}}

%--right input path
\put(145,310){\line(1,-1){130}}
%--right beam splitter
\put(255,175){\line(0,1){50}}
%--right detector
\put(267,172){\line(1,1){16}}
\bezier{80}(267,172)(277,162)(285,170)
\bezier{80}(285,170)(293,178)(283,188)
\bezier{80}(285,170)(295,160)(305,170)
\put(285,170){\makebox(40,20){$\langle 1\!\mid$}}

%--left input path
\put(255,310){\line(-1,-1){130}}
%--left beam splitter
\put(145,175){\line(0,1){50}}
%--left detector
\put(133,172){\line(-1,1){16}}
\bezier{80}(133,172)(123,162)(115,170)
\bezier{80}(115,170)(107,178)(117,188)
\bezier{80}(115,170)(105,160)(95,170)
\put(75,170){\makebox(40,20){$\langle 1 \!\mid$}}

%--right output path
\put(270,215){\line(-1,-1){125}}
%--right photon source
\put(270,215){\line(1,0){17}}
\put(270,215){\line(0,1){17}}
\put(289,250){\line(-1,-1){18}}
\put(305,234){\line(-1,-1){18}}
\put(305,234){\line(-1,1){16}}
\put(278,221){\makebox(18,18){$\mid\! 1 \rangle$}}

%--left output path
\put(130,215){\line(1,-1){95}}
\put(245,100){\line(1,-1){10}}
%--left photon source
\put(130,215){\line(-1,0){17}}
\put(130,215){\line(0,1){17}}
\put(111,250){\line(1,-1){18}}
\put(95,234){\line(1,-1){18}}
\put(95,234){\line(1,1){16}}
\put(106,221){\makebox(18,18){$\mid\! 1 \rangle$}}

%==VV-attenuation
%--3/4 beam splitter 
\put(95,95){\line(0,1){50}}
\put(100,120){\makebox(40,20){$R=3/4$}}
%--subtracted line & detector
\put(95,120){\line(-1,-1){15}}
\put(88,97){\line(-1,1){16}}
\bezier{80}(88,97)(78,87)(70,95)
\bezier{80}(70,95)(62,103)(72,113)
\bezier{80}(70,95)(60,85)(50,95)
\put(30,95){\makebox(40,20){$\langle 0 \!\mid$}} 

%==phase correction
\put(230,125){\line(1,-1){20}}
\put(220,115){\line(1,-1){20}}
\put(230,125){\line(-1,-1){10}}
\put(250,105){\line(-1,-1){10}}
\put(230,110){\makebox(60,20){$\Delta \phi = \pi$}}

\end{picture}

\caption{\label{setup} Schematic setup of the quantum filter
for two photon polarization correlations. Vertical lines
represent beam splitters with a reflectivity of $R=1/2$, 
unless labled otherwise. The boxes labled H|V represent
polarization sensitive beam splitters transmitting H polarized
photons and reflecting V polarized ones. Quantum states created
by single photon sources are given by bra-notation, and 
post-selected detector states by ket-notation. Note that 
the $R=3/4$ beam splitter has a vacuum input not explicitly
given in the figure.}
\end{figure}

Figure \ref{setup} shows the complete setup for the two photon
polarization filter. The filter described by equation 
(\ref{eq:core}) is applied to the horizontally polarized 
components of the two input photons. 
The $\mid\! 1;1\rangle$ component therefore
corresponds to the $\mid\! H;H \rangle$ component of the
two photon state, and the $\mid\! 0;0\rangle$ component
corresponds to the $\mid\! V;V \rangle$ component.
The attenuation of the $\mid\! V;V \rangle$ component
is realized by the beam splitter of reflectivity $R=3/4$ in
one of the vertically polarized paths. A photon detector
post-selects only events with zero reflected photons, 
effectively suppressing the one photon component by an 
amplitude factor of $1/2$. The components of
$\mid\! V;H \rangle$ and $\mid\! H;V \rangle$ correspond
to the $\mid\! 0;1 \rangle$ and $\mid\! 1;0 \rangle$
states in equation (\ref{eq:core}) and are therefore 
completely suppressed by the filter. The operation of 
this filter can now be summarized in a single operator
$\hat{S}_{\mbox{total}}$ describing the effect on an
arbitrary two photon input state. 
It reads
\begin{equation}
\label{eq:filter}
\hat{S}_{\mbox{total}} = \frac{1}{4} \left(
\mid\! H;H\rangle\langle H;H\!\mid +  
\mid\! V;V\rangle\langle V;V\!\mid \right),
\end{equation}
This operator projects the input state onto the two dimensional
subspace of identical horizontal or vertical polarization,
where the factor of $1/4$ is an expression of the post-selection
efficiency indicating that, even within the transmitted 
subspace of the parallel polarization components 
$\mid H;H \rangle$ and $\mid V;V \rangle$, 
only $1/16$ of the input states pass the filter.

The most significant property of this filtering process 
is that it leaves the entanglement properties of the input
intact by preserving the quantum coherence between 
$\mid\! H;H\rangle$ and $\mid\! V;V\rangle$. In fact, it can 
be shown that any maximally entangled photon pair will still 
be maximally entangled if it passes this filter. 
In the H,V representation, all maximally entangled 
states are of the form
\begin{equation}
\mid \mbox{max.}(c_1,c_2,\phi)\rangle = \frac{1}{\sqrt{2}} 
\left(\mid\! H\rangle \otimes 
(c_1 \mid\! H\rangle + c_2 \mid\! V\rangle) +
 e^{-i\phi} \mid\! V\rangle \otimes 
(c_2 \mid\! H\rangle - c_1 \mid\! V\rangle) \right). 
\end{equation}
The application of the filter to an entangled photon pair of
this kind then produces an output of
\begin{equation}
\hat{S}_{\mbox{total}} \mid \mbox{max.}(c_1,c_2,\phi)\rangle = 
\frac{c_1}{4 \sqrt{2}} 
\left(\mid\! H;H\rangle -
 e^{-i\phi} \mid\! V;V \rangle \right). 
\end{equation}
This property demonstrates that the filter is only sensitive to 
non-local polarization correlations of the two photons. It is 
therefore a suitable tool for entanglement manipulations.

The clearest physical description of the 
filter properties is that of a quantum nondemolition measurement  
for the correlation of the H-V polarization of the photon pair.
In this context, it is remarkable that the relative orientations
of all other polarizations remain unchanged, even though the
local polarizations are necessarily randomized by the measurement
back-action. For example, it is possible to put one right
and one left circular polarized photon through the filter. 
The output will still have opposite circular polarization,
while all linear polarizations will be parallel,
\begin{eqnarray}
\label{eq:entangle}
\hat{S}_{\mbox{total}}\mid\! R;L\rangle &=&
\hat{S}_{\mbox{total}} \;
\frac{1}{\sqrt{2}} 
\left(\mid\! H \rangle + i \mid\! V \rangle \right)
\otimes \frac{1}{\sqrt{2}} 
\left(\mid\! H \rangle - i \mid\! V \rangle \right)
\nonumber \\ &=& 
\frac{1}{8}(\mid\! H;H \rangle + \mid\! V;V \rangle)
\hspace{0.2cm}
= \frac{1}{8}(\mid\! R;L \rangle + \mid\! L;R \rangle).
\end{eqnarray}
This filter can thus be used to entangle two previously 
independent photons. 
Moreover, the filter can create multi-photon entanglement
if it is applied to members of entangled pairs, for example
to photon 2 of an entangled pair (1,2), and to photon 3 of
an entangled pair (3.4). The output of this filtering process
then reads
\begin{eqnarray}
\hat{S}_{\mbox{total}}(2,3) \;
\frac{1}{\sqrt{2}} 
\left(\mid\! H;H \rangle_{1,2} + \mid\! V;V \rangle_{1,2} \right)
\otimes \frac{1}{\sqrt{2}} 
\left(\mid\! H;H \rangle_{3,4} + \mid\! V;V \rangle_{3,4} \right) 
&=&
\nonumber \\ && \hspace{-4cm}
\frac{1}{8}(\mid\! H;H;H;H \rangle + \mid\! V;V;V;V \rangle).
\end{eqnarray}
While this method of generating multi-photon entanglement
may seem a bit more complicated than the simple post-selection
schemes recently realized to generate three- and four-photon
entanglement \cite{Bou99,Pan01}, it has the advantage of allowing
only single photon states in the output ports. Therefore, it
can be used as a reliable source of multi-photon entanglement,
even in applications where post-selection of one photon events 
in the output ports is not an option.

Another important feature of the filter is the transfer of
quantum coherence from single photon states to multi-photon
states. If one of the input photons is polarized along the 
diagonal between the H and V directions, the filter transfers
the quantum coherence of the other input photon to a two
photon state,
\begin{eqnarray}
\hat{S}_{\mbox{total}} \;
\frac{1}{\sqrt{2}} 
\left(\mid\! H \rangle + \mid\! V \rangle \right)
\otimes
\left(c_H \mid\! H \rangle + c_V \mid\! V \rangle \right) &=&
\nonumber \\ && \hspace{-4cm}
\frac{1}{4 \sqrt{2}}(c_H \mid\! H;H \rangle 
+ c_V \mid\! V;V \rangle).
\end{eqnarray}
This process conserves the quantum information carried by the
second photon. Effectively, the output represents a very
simple form of a two photon quantum code. Such a quantum 
code could be useful to improve the error resistance of 
quantum computation and teleportation \cite{Kni01,Sho95}.
In particular, this code makes the qubit resistant against 
decoherence caused by projective measurements of the circular 
polarization or of polarization components along the diagonals 
between the H and V directions. Such errors are typical for
the basic implementation of quantum teleportation using
polarization entanglement \cite{Bou97}. Therefore, this
type of code may be useful for the enhancement of 
teleportation efficiency. 
The transfer of quantum information from a single photon
qubit to a multi-photon state may be generalized to an
arbitrary number of photons by using entangled input
photons. For example, quantum coherence is transferred
to a three photon state by
\begin{eqnarray}
\hat{S}_{\mbox{total}}(2,3) \;
\frac{1}{\sqrt{2}} 
\left(\mid\! H;H \rangle_{1,2} + \mid\! V;V \rangle_{1,2} \right)
\otimes
\left(c_H \mid\! H \rangle_3 + c_V \mid\! V \rangle_3 \right) &=&
\nonumber \\ && \hspace{-4cm} 
\frac{1}{4 \sqrt{2}}(c_H \mid\! H;H;H \rangle 
+ c_V \mid\! V;V;V \rangle).
\end{eqnarray}
In general, filtering one photon of an N-1 photon entangled 
state and the original single photon qubit automatically 
encodes the input qubit into a corresponding N-photon code.

The implementation of quantum codes and the creation of
entanglement between previously unentangled inputs are
just two examples for a wide range of applications.
As can be seen from equation (\ref{eq:filter}), the quantum 
filter described above realizes an elementary form of 
entanglement manipulation. Moreover, the setup 
shown in figure \ref{setup} can be realized using only
standard elements of linear optics and photon detectors. 
However, it should be
noted that a high level of precision is required to avoid
errors in the filtering process.
For instance, it is necessary to avoid dark counts
in the detectors. This can be achieved by limiting
the time window of the photon detection to the pulse length of
the single photon inputs. For example, at a pulse length of one 
nanosecond, the dark count rate of 10000 counts per second 
estimated for the detectors presented in \cite{Tak99,Kim99} 
translates into a mere $10^{-5}$ counts per pulse.
Also, care should be taken to optimize mode matching and pulse
timing to achieve maximal photon bunching in the beam splitters.
This requirement is typical for experiments using photon 
entanglement, and previous experimental results suggest that 
sufficient precision can be achieved using conventional 
techniques \cite{Bou99,Pan01,Bou97}. 
A potentially more serious difficulty 
arises from limited detector efficiencies. The detectors in the H 
polarized paths need to distinguish between one photon 
events and two photon events. While the detectors presented 
in \cite{Tak99,Kim99} 
should be well suited for this task, the quantum efficiency 
for this operating regime of the detectors is presently 
limited to $88\%$. This means that there is a $19\%$ chance 
of mistakenly registering a two photon event as a one 
photon event. 
Unfortunately, a detailed analysis of this error is beyond 
the scope of this letter. However, a good estimate of the 
error frequency can be obtained by focusing on the errors 
for the $\mid\! H;V \rangle$ input state. For the setup given in
figure \ref{setup}, this input state has the highest error 
frequency, since the $\mid\! H;V \rangle$ component is 
filtered out entirely by zero or two photon event 
at the detectors in the H polarized paths. 
The probability of a two photon event at one of the detectors
and a one photon event at the other is $1/4$. Therefore, the 
total frequency of errors at a quantum efficiency of 
$88\%$ is given by $19\%\,/4 \approx 5 \%$.
While this error rate is still very low, it must be 
compared to the filter efficiency of $1/16 = 6.25 \%$ for 
$\mid\! H;H \rangle$ and $\mid\! V;V \rangle$ input states.
For example, the operation described in equation 
(\ref{eq:entangle}) will produce the correct entangled state
with a probability of $3.13 \%$. However, the chance of a 
mistaken transmission due to the $25 \%$ contribution of
the $\mid\! H;V \rangle$ component in the input is about $1.25\%$.
This mistake would then lead to a mixture of about
$70\%$ entangled photon pairs originating from successful filter 
operations, and about $30\%$ V polarized single photon outputs 
originating from failure to detect both of two H polarized photons
at one of the detectors. This example illustrates the importance
of further increasing detector efficiency to distinguish
one and two photon events with greater reliability. 

In conclusion, a quantum filter for photonic qubit pairs can be
realized using only linear optics and single photon counters.
Such a filter may be a very useful component for optical 
quantum information networks. We have presented various 
applications and discussed the technological
requirements for an experimental realization of this scheme. 
At present, the major technological challenge appears to be 
the improvement of detector efficiency to distinguish single
photon events from two photon events. If this difficulty can
be overcome, the quantum filter for non-local polarization 
properties of photonic qubits may proof to be a very flexible 
tool for the implementation of photon based quantum information 
technologies.

\end{document}